**Decoding hand kinematics from population responses in sensorimotor cortex during grasping**

Elizaveta V. Okorokova[1,2], James M. Goodman[1], Nicholas G. Hatsopoulos[1,3] & Sliman J. Bensmaia[1,3]

[1]Committee on Computational Neuroscience, University of Chicago, Chicago, IL

[2]Department of Psychology, Higher School or Economics, Moscow, Russia

[3]Department of Organismal Biology and Anatomy, University of Chicago, Chicago, IL

**Acknowledgements**

We thank Sangwook A. Lee, Gregg A. Tabot, Alexander T. Rajan for help with data collection. This work was supported by NINDS R01 NS 082865.

**ABSTRACT**

The hand – a complex effector comprising dozens of degrees of freedom of movement – endows us with the ability to flexibly, precisely, and effortlessly interact with objects. The neural signals associated with dexterous hand movements in primary motor cortex (M1) and somatosensory cortex (SC) have received comparatively less attention than have those that are associated with proximal limb control. To fill this gap, we trained three monkeys to grasp objects varying in size, shape, and orientation while tracking their hand postures and recording single-unit activity from M1 and SC. We then decoded their hand kinematics across 30 joints from population activity in these areas. We found that we could accurately decode kinematics with a small number of neural signals and that performance was higher for decoding joint angles than joint angular velocities, in contrast to what has been found with proximal limb decoders. We conclude that cortical signals can be used for dexterous hand control in brain machine interface applications and that postural representations in SC may be exploited via intracortical stimulation to close the sensorimotor loop.

**INTRODUCTION**

The hand, a complex effector comprising dozens of degrees of freedom (Belic & Faisal, 2015), allows us to flexibly, precisely, and effortlessly manipulate objects. The loss of hand function – as a consequence of spinal cord injury, for example – can have devastating consequences on quality of life (Anderson, 2004). In patients whose sensorimotor cortex is intact, some measure of independence can be restored with brain-machine interfaces (BMIs) that tap into the central neural pathways mediating manual dexterity (Bensmaia & Miller, 2014). Translating patterns of neural activity into outputs of external devices is critical to advance such interfaces.

Reach-to-grasp movements have been traditionally decoded from cortical areas associated with the planning and execution of movement. Primary motor, premotor, and posterior parietal cortices have been the main targets of extant BMIs, yielding remarkable control of a robotic limb in both non-human primates (Lebedev et al., 2005; Mulliken, Musallam, & Andersen, 2009; Velliste, Perel, Spalding, Whitford, & Schwartz, 2008; Wessberg et al., 2000) and human tetraplegic patients (Hochberg et al., 2012; S.-P. Kim, Simeral, Hochberg, Donoghue, & Black, 2008; S. Kim et al., 2011; Wodlinger et al., 2015). However, the focus of most decoding studies has been on the proximal limb (elbow and shoulder) performing reaching movements (for example, Dyer et al., 2017; Gilja et al., 2012). The distal limb (wrist and finger joints), critical to object interactions, has received comparatively little attention (but see Menz, Schaffelhofer, & Scherberger, 2015), primarily due to the difficulty of simultaneously tracking tens of hand joints (S. Schaffelhofer & Scherberger, 2012) and the greater complexity of manual behavior (Ingram, Körding, Howard, & Wolpert, 2008).

A critical complement to neural signals involved in controlling movements are signals responsible for conveying sensory feedback about the consequences of those movements (Scott, 2004). Indeed, the motor apparatus receives continuous proprioceptive feedback – primarily from muscles – that tracks the position of the body in space and the forces it exerts, and mediates error-corrective motor adjustments (Soechting & Flanders, 1989). Proprioceptive impairments lead to major deficits in motor behavior, leading to slow, imprecise, and effortful movements (Cole & Sedgwick, 1992; Ghez & Salnbosrg, 1995; Sainburg, Ghilardi, Poizner, & Ghez, 1995). Two cortical fields in somatosensory cortex (SC) – Brodmann's areas 3a and 2 (Krubitzer, Huffman, Disbrow, & Recanzone, 2004; Pons & Kaas, 1986) – contain neurons that respond during active movements, passive manipulation of joints and muscles (Gardner & Costanzo,

1981; B. M. London & Miller, 2013; Prud'homme & Kalaska, 1994), and when forces are applied to the joints (Fromm & Evarts, 1982). However, the vast majority of extant studies of proprioceptive representations in SC have focused on the proximal limb, particularly during reaching movements.

Of the existing research on representations of hand movements in SC, two lines of research have investigated the responses of *individual* SC neurons, one during passive deflections of the hand joints (Costanzo & Gardner, 1981; Gardner & Costanzo, 1981) and the other during active hand movements (Goodman et al., 2019). However, the degree to which hand movements and postures are encoded across *populations* of SC neurons has not been investigated. One way to address this question is to assess our ability to decode hand kinematics from the responses of populations of SC neurons. Earlier studies have attempted to decode kinematics using SC activity; however, they mainly focused on tracking a finite number of discrete hand configurations (Branco et al., 2017; Farrokhi & Erfanian, 2018) or only considered tasks such as reach-to-grasp that combined simple grasping behavior with comparatively rich reaching kinematics (Carmena et al., 2003; Glaser, Chowdhury, Perich, Miller, & Kording, 2017). To the extent that somatosensory neurons can be established to carry detailed information about hand movements, these neural representations might be exploited to convey artificial proprioceptive feedback through intracortical stimulation (ICMS)(Brian M. London, Jordan, Jackson, & Miller, 2008; Salas et al., 2018; Tomlinson & Miller, 2016).

The goal of the present study is to assess the degree to which hand kinematics can be decoded from the responses of populations of sensorimotor neurons. To this end, we trained three monkeys to grasp 35 objects of varying sizes, shapes, and orientations, while tracking their time-varying hand kinematics using a camera-based motion tracking system and simultaneously recording the responses in the hand representations in M1 and SC using chronically implanted electrode arrays. We show that we can decode the postures and movements of 30 joints of the monkey's hand as it preshapes to grasp each object with as few as 20 neurons in each area. Furthermore, we find that we can better decode posture than movement, in contrast to what has been shown for motor representations of the proximal limb, suggesting possible differences in cortical encoding for proximal and distal limb. Our results underscore the promise of using M1 signals to achieve dexterous control of the hand and demonstrate that SC populations also carry a faithful representation of time-varying hand configuration that might be exploited to restore proprioception through ICMS.

## METHODS

### Animals and surgery

We recorded from three male Rhesus macaques ranging in age between 6-15 years and weighing between 8 and 11 kg. All animal procedures were performed in accordance with the rules and regulations of the University of Chicago Animal Care and Use Committee. Monkeys received care from a full-time husbandry staff who maintained a 12hr/12hr light/dark cycle, cleaned the animals' living spaces once a week, and provided the animals with ample food and enrichment. In addition, a full-time veterinary staff monitored the animals' health. The water intake of the animals was regulated according to a protocol requiring monitoring their weights daily and ensuring a minimum daily water consumption of 10 cc/kg.

Surgical procedures consisted of implantation of a head-fixing post onto the skull, craniotomy, implantation of a sealed recording chamber and chronic recording arrays. Monkey 1 was implanted with two Utah electrode arrays (UEAs, Blackrock Microsystems, Inc., Salt Lake City, UT), one in primary motor cortex, the other in somatosensory cortex and four floating microelectrode arrays (FMAs, Microprobes for Life Science, Gaithersburg, MD), two in the anterior and two in the posterior bank of the central sulcus (Figure 1D). Monkeys 2 and 3 were implanted with semichronic Microdrive electrode arrays (SC96, Gray Matter Research, Bozeman, MT), each spanning large swaths of primary motor and somatosensory cortex and comprising individually depth-adjustable electrodes (**Error! Reference source not found.**D) (Figure 1E) (Dotson et al., 2017). All procedures were performed under aseptic conditions and under anesthesia induced with ketamine HCl (20 mg/kg, IM) and maintained with isoflurane (10-25 mg/kg per hour, inhaled).

*Behavioral task*

We trained monkeys to grasp 35 objects varying in shape, size and orientation (Figure 1A, inset). Throughout the session, the head-fixed monkey sat in a chair facing a three DoF robotic arm (Figure 1A, top). The hand of the monkey rested on the cushioned armrest equipped with a photo sensor to ensure that the forearm was largely immobile during the grasp. At the beginning of each trial, an object was attached to the robotic arm using a weak magnet and presented to the animal. The monkey's task was to grasp the object and exert enough grip force so that, when the robot retracted, the object would be disengaged from its magnetic coupling with the robot and remain in the monkey's hand (Figure 1B). Animals were trained to keep their elbow on the armrest to minimize movements of the proximal limb.

*Kinematics*

We recorded hand and elbow kinematics using a camera-based motion tracking system (Vantage, VICON, Los Angeles, CA). To this end, we placed 30 reflective markers on the joints of the hand, wrist, and proximal limb (Figure 1C, Figure 2 top). Ten cameras were used to capture the kinematics of the first monkey at a rate of 250 Hz, and fourteen cameras were used to capture the kinematics of the other monkeys at a rate of 100 Hz. We then labeled each marker using Nexus software (VICON, Los Angeles, CA) and performed inverse kinematic modeling (OpenSim, Scott L. Delp et al., 2007) from the resulting time-varying marker positions to infer time-varying joint angles of the limb (Table 2). Joint angles were smoothed with a 50-ms moving average and the angular velocities were computed from these. For each trial, we identified the start of movement, maximum aperture of fingers, and contact with an object. We labeled these events manually for a subset of trials for each animal and then used linear discriminant analysis to infer maximally likely event times for other trials. The joint angles we traced included: elbow flexion, wrist pronation/supination, wrist deviation, wrist flexion/extension, 1 mcp flexion, 1 cmc supination, 1 cmc abduction, 1 mcp flexion, 1 mcp abduction, 1 ip flexion, 2 mcp flexion, 2 mcp abduction, 2 pm flexion, 2 md flexion, 3 mcp flexion, 3 mcp abduction, 3 pm flexion, 3 md flexion, 4 mcp flexion, 4 mcp abduction, 4 pm flexion, 4 md flexion, 4 cmc extension, 5 mcp flexion, 5 mcp abduction, 5 pm flexion, 5 md flexion, 5 cmc extension, 5 cmc abduction, 5 cmc supination.

*Electrophysiology*

We recorded neural signals from monkey 1 using UEAs placed in the post- and pre-central gyri and from FMAs placed in the posterior and anterior banks of the central sulcus (Figure 1D). For monkeys 2 and 3, we recorded neural signals using arrays of depth-adjustable electrodes (SC96) positioned over the central sulcus (Figure 1E, Supplementary Figure 1). We used offline spike sorting (Offline Sorter, Plexon, Dallas, TX) to remove non-spike threshold crossings and to isolate individual units when more than one was present on a given trace. For comparison between areas, we divided M1 neurons into caudal and rostral

M1 and SC neurons into area 2 and 3a based on the depth of electrodes and receptive fields. Histological reconstructions, obtained for one monkey (Monkey 3), verified the location of proprioceptive neurons in Brodmann's areas 3a and 2 (see Goodman et.al. 2019 for details).

*Neural data preprocessing*

Time-varying firing rates of all recorded neurons were computed by summing all events in 10-ms bins. We then soft-normalized the firing rates (divided by their range plus a small increment) and convolved the resulting rates with a Gaussian kernel with 100-ms width (Figure 2, bottom). We then reduced the dimensionality of the neural space with Principal Component Analysis (PCA) and preserved components that cumulatively explained at least 90% of the variance in the neural data.

Session stitching

To achieve a sufficient sample size from each cortical field, we pooled data from all sessions for each monkey (see Table 1). To pool kinematics across sessions, we aligned the kinematics from the same condition (object) to maximum hand aperture and averaged them across sessions for each monkey separately (Supplementary Figure 2). To eliminate trials on which the animal used a different grasping strategy for a given object, we discarded trials on which the worst joint correlation between that trial's kinematics and the mean kinematics was below 0.7. We then used the average kinematic traces 600 ms before and 200 ms after the alignment point for decoding (all before object contact).

Decoding

To predict hand and arm kinematics, we applied the Kalman filter (Kalman 1960), commonly used for kinematic decoding (Wu et al. 2004; Menz et al. 2015). In this approach, kinematic dynamics can be described by a linear relationship between past and future states:

$$x_t = Ax_{t-1} + v_t \qquad (1)$$

where $x_t$ is a vector of joint angles or joint velocities at time $t$, $A$ is a state transition matrix, and $v_t$ is a vector of random numbers drawn from a Gaussian distribution with zero mean and covariance matrix $V$. The kinematics $x_t$ can be also explained in terms of the observed neural activity $z_{t-\Delta}$:

$$x_t = Bz_{t-\Delta} + w_t \qquad (2)$$

Here, $z_{t-\Delta}$ is a vector of instantaneous firing rates across a population of M1 neurons at time $t - \Delta$, $B$ is an observation model matrix, and $w_t$ is a random vector drawn from a Gaussian distribution with zero mean and covariance matrix $W$. We tested multiple values of the latency, $\Delta$, and report decoders using the latency that maximized decoder accuracy (150ms, Supplementary Figure 3).

We estimated the matrices $A, B, V, W$ using linear regression on each training set, and then used those estimates in the Kalman filter update algorithm to infer kinematics of each corresponding test set (see Faragher 2012; Okorokova et al. 2015 for details). Briefly, at each time $t$, kinematics were first predicted using the state transition equation (1), then updated with observation information from equation (2). Update of the kinematic prediction was achieved by a weighted average of the two estimates from (1) and (2): the weight of each estimate was inversely proportional to its uncertainty (determined in part by $V$ and $W$ for the estimates based on $x_{t-1}$ and $z_{t-\Delta}$, respectively), which changed as a function of time and was thus recomputed for every time step.

We performed 10-fold cross-validation, in which we trained the parameters of the filter ($A, B, V, W$) on a randomly selected 90% of the trials and tested the model using the remaining 10% of trials. If the number of neurons differed between areas, we randomly picked the same number of units from each area, repeating the procedure 10 times. Performance was assessed using the coefficient of determination ($R^2$). To find the optimal lag, we tested the model at various lags and selected the one that yielded the best cross-validated performance (150ms, Supplementary Figure 3).

To compare our result to previously reported metrics, we additionally computed Pearson's correlation and the normalized Root Mean Squared Error (rRMSE), which is RMSE divided by the range of each respective degree of freedom (i.e. average error as a percentage of the joint's range of motion, see Menz et.al. (2015) for details). We compared the standard Kalman filter decoder to other types of linear and non-linear decoders, including Wiener Filter (WF), Wiener Cascade Filter (WFC), Extreme Gradient Boosting (XGBoost), Dense Feedforward Neural Network (DNN), Recursive Neural Network (RNN), Gated Recurrent Unit (GRU) and Long Short Term Memory Network (LSTM) described in detail in Glaser et al. (2017).

**RESULTS**

*Decoding kinematics averaged across trials*

First, we assessed the degree to which the responses of neuronal populations in M1 and SC convey information about time-varying hand kinematics, pooled across multiple recording sessions (Figure 3, Supplementary Figure 4). We found that, from a population of randomly selected neurons (N=20), we could reconstruct time-varying joint angles accurately for most joints, obtaining a median performance ($R^2$) of 0.62, 0.66, and 0.63 for M1, SC, and the two combined, respectively (Figure 3B). For 2 out of 3 monkeys, M1 decoders outperformed, on average, those based on SC (Figure 3C). However, for monkey 2, the reverse was true. This difference is likely due to the respective locations of SC recordings: In monkeys 1 and 3, the majority of SC neurons were located in area 2, whereas in monkey 2, they were in area 3a (see Supplementary Table 1). Decoding was only marginally improved with more recently developed decoders, including non-linear ones (Supplementary Figure 5).

*Comparing cortical areas*

Next, we performed a more detailed analysis of how information about hand posture is distributed within M1 and SC (Figure 4). Indeed, the caudal region of M1 contains more corticomotor (CM) neurons – which make monosynaptic connections with motoneurons – than does its rostral counterpart, and these CM neurons are thought to be critical for highly skilled movements, particularly of the hand (Rathelot & Strick, 2009). Furthermore, while neurons in Brodmann's area 3a exhibit almost exclusively proprioceptive responses, those in area 2 exhibit mixed proprioceptive and cutaneous responses (S. S. Kim, Gomez-Ramirez, Thakur, & Hsiao, 2015), and the latter may obscure kinematic signals. Decoding accuracy depends on the size of the neuronal sample, so we measured performance as a function of sample size for all populations (Figure 4A). We did not find any systematic differences between caudal and rostral M1. However, decoding based on area 2 was systematically worse than that based on other cortical fields. In the one animal with a sufficient sample size, decoding from area 3a achieved better performance than from M1. With perhaps the exception of area 2, then, the various cortical fields seem to contain similar amounts of information about hand kinematics.

*Decoding the kinematics of groups of joints*

Next, we investigated whether the M1 and SC subpopulations that were sampled preferentially encoded movements of different portions of the limb. To this end, we divided the joints into 7 groups – proximal arm (elbow), wrist (pronation-supination, flexion, extension), and the five digits (separately) – and assessed the average decoding performance for each group. We did not find any systematic patterns, suggesting that the arrays impinged upon neural populations that spanned the hand and arm representation in each area (Supplementary Figure 6).

*Decoding postural synergies*

Joint kinematics of the hand have been shown to exhibit systematic correlational structure, with some joints tending to move together. A common way to reveal this structure is through principal components analysis, which yields a set of mutually orthogonal bases of kinematics (Figure 5A). We examined the degree to which we could decode each of the principal components from the neuronal responses. When measured using the coefficient of determination, we found that decoder performance dropped sharply for the lower-variance components, with only the first three components showing performance comparable to that for single joints (Figure 5B-C). However, the drop in performance for the low-variance components vanishes when model fit is expressed using a metric, rRMSE, that is less sensitive to signal variance.

*Decoding postures vs. movements*

In all of the above analyses, we showed that time-varying postures could be directly decoded from neuronal activity. This approach stands in contrast to that adopted for proximal limb kinematics or the application of proximal limb-related M1 activity to cursor control, which typically involves decoding joint or endpoint velocities from neuronal responses and then integrating these to obtain postures (Chase, Schwartz, & Kass, 2009; S.-P. Kim et al., 2008; Koyama et al., 2010; Taylor, Tillery, & Schwartz, 2002; Velliste et al., 2008). Indeed, even SC decoders for reaching movements appear to perform better when decoding limb velocity than when decoding limb posture (Weber et al., 2011). With this in mind, we wished to directly test whether neuronal responses in M1 and SC preferentially encode postures or movements during grasp. To this end, we reconstructed joint angular velocities from sensorimotor responses and compared these to our reconstructions of angular positions. For this analysis, we used a single lag – at 150 ms determined to yield peak performance (Supplementary Figure 3) – because multi-lag models allow for integration (from a velocity to a position signal) or differentiation (from a position to a velocity signal) so obscures the distinction between postural and movement coding. Using 20 units from each cortical subpopulation, we found that postures could be significantly better reconstructed than could movements ($p<0.01$, paired t-tests for each area and monkey), in contrast to what has been observed for the proximal limb (Figure 6).

*Single trial decoding*

Finally, we gauged the degree to which we could decode single-trial kinematics from sensorimotor cortex and verified that results described above were not artifacts of pooling data across sessions. To this end, we analyzed data from one animal, from whom a sufficient sample size had been achieved on a pair of single sessions ($N_1 = 44$, $N_2 = 36$ from M1 and $N_1 = 15$, $N_2 = 9$ from S1). Specifically, we decoded joint angles and velocities obtained on single trials from the concurrently recorded neuronal responses and compared the resulting performance to that of the average kinematic decoders for the same joints and velocities. We found that decoding performance of pooled and single trial kinematics was equivalent ($p>0.05$, Figure

7A,B), which may at first be surprising. Indeed, one might expect pooling to degrade performance to the extent that the kinematics are not identical. However, our results suggest that kinematics are similar within monkeys across sessions (Supplementary Figure 2). The deterioration of decoding of pooled time-varying hand kinematics decoding due to differences in kinematics across trials is offset by the noisier neuronal signals used for single-trial decoding. We also verified that postural decoders significantly outperformed movement decoders even for single trial responses (Figure 7C).

## DISCUSSION

### High dimensional decoding

Most decoding studies to date focused on continuous movement of a few degrees of freedom, usually including shoulder and elbow (Ganguly & Carmena, 2009; Gilja et al., 2012; Lebedev et al., 2005; Mulliken, Musallam, & Andersen, 2009; Suminski, Tkach, Fagg, & Hatsopoulos, 2010; Wessberg et al., 2000; Yu et al., 2007 and others) and, less frequently, wrist (Hochberg et al., 2012; Wodlinger et al., 2015) and finger joints (Aggarwal et al., 2013; Menz et al., 2015). Because of the complexity of the space of hand kinematics, most previous efforts to decode hand postures were either discrete, focusing on classification of a finite number of finger-wrist configurations (Branco et al., 2017; Carpaneto et al., 2011; Chestek et al., 2013; Stefan Schaffelhofer, Agudelo-toro, & Scherberger, 2015) or limited to a few common continuous finger movements, such as pinch, scoop, grip and whole finger flexion/extension (Acharya et al., 2009; Bansal, Truccolo, Vargas-irwin, & Donoghue, 2012; Hochberg et al., 2012; Wodlinger et al., 2015). As hand tracking technologies become increasingly available (Mathis et al., 2018; Pereira et al., 2019; S. Schaffelhofer & Scherberger, 2012), decoding tens of joints simultaneously will become increasingly manageable (see for example, Menz et al., 2015). Here, we show that up to 30 joints of the upper extremity can be decoded with a relatively small population of sensorimotor neurons even with a simple linear decoder.

### Decoding from M1

Our approach is similar to that described in Menz et al. (2015), in which 27 degrees of freedom of grasping kinematics were reconstructed from the responses of neurons in posterior parietal cortex and M1. Decoders built from the one area common to both studies (M1) yielded comparable performance (Supplementary Figure 4).

Anatomically, M1 can be divided into rostral and caudal regions. Neurons in the caudal region of M1 have direct connections with motoneurons in the spinal cord whereas neurons in the rostral region contact mainly spinal interneurons and thus drive muscles only indirectly (Rathelot & Strick, 2009). Neurons in caudal M1 might be particularly relevant for dexterous hand control as evidenced by the cross-species correlation between manual dexterity and the preponderance of cortico-motor neurons (Bortoff & Strick, 1993; Maier et al., 1997). Despite the documented anatomical differences, we did not see any differences in decoding performance between the two areas in monkeys for which both M1 populations were available. As grasp comprises highly correlated patterns of joint movements, this manual behavior may not require a direct line to muscles. Caudal and rostral M1 may be differentially engaged in non-prehensile dexterous movements that involve more individuated finger movements (Bortoff & Strick, 1993; Maier et al., 1997).

### Decoding from SC

Previous attempts to decode kinematics from SC activity focused on proximal limb movements. Adding signals from SC significantly improved performance in macaques engaged in reaching beyond the performance achieved with M1 signals alone (Carmena et al., 2003; Lebedev et al., 2005). Decoding limb kinematics from electrocorticographic (ECoG) signals in SC achieves similar performance as with ECoG signals in M1 (Branco et al., 2017; Farrokhi & Erfanian, 2018).

In the present study, we decode, for the first time, hand kinematics from the spiking activity of neurons in areas 3a and 2 and find that both areas yield well above chance performance in all joints. Area 3a showed performance comparable to M1, consistent with earlier observations of single-unit responses (Goodman et al., 2019). Area 2, which lies downstream of area 3a, yielded significantly worse performance than M1, consistent with the observation that this cortical field also receives substantial cutaneous input, which may obscure the proprioceptive representations, even before contact. While movement has been previously shown to activate cutaneous neurons in the absence of contact (S. S. Kim et al., 2015; Rincon-gonzalez, Warren, Meller, & Tillery, 2011), our results suggest that these cutaneous responses may not support kinematic encoding of the hand.

*Posture and movement decoding*

Neurons in motor cortex and proprioceptive areas of somatosensory cortex have been shown to preferentially encode velocity of the proximal limb (movement), rather than its position (posture) (Paninski, Fellows, Hatsopoulos, & Donoghue, 2004; Wang, Chan, Heldman, & Moran, 2007). Consistent with this finding, decoders of joint velocities generally outperform those of joint positions (S.-P. Kim et al., 2008). However, this preferential encoding and decoding of movement over posture has been tested exclusively for proximal limb joints. When we directly compared posture vs. movement decoding of the hand, we found that posture can be more faithfully decoded than can movement (Figure 6) and this postural preference is not an artefact of averaging (Figure 6C). For this analysis, we restricted the decoder to a single lag to avoid the effect of linear integration or differentiation that would confound the result. Indeed, in multiple-lag model, the difference in performance between posture and movement decoders becomes less pronounced (Supplementary Figure 7). The postural preference for hand-related sensorimotor neurons implies a difference between proximal and distal limb representations, which may be inherited from the different inertial and biomechanical properties of the arm and the hand. A neuronal representation of the hand that emphasizes posture is well suited to support stereognosis (Goodman et al., 2019).

*Decoding methods*

The application of machine learning to kinematic/cursor decoding is now standard practice (Glaser et al., 2017). However, the extent to which recently developed decoding approaches robustly improve performance of high dimensional decoders (of hand kinematics, e.g.) has not been investigated. Here, we applied a variety of well-established linear and non-linear approaches to decoding hand movements (described in detail in Glaser et al., 2017) and found that non-linear methods generally outperform standard linear ones, as might be expected (Supplementary Figure 5). However, the improvement is typically minimal and may not justify the added computational complexity and potential for overfitting.

*Closed-loop robotic limb control*

A major application of kinematic decoders is to drive brain machine interfaces aimed at restoring movement in patients with sensorimotor impairments (Bensmaia & Miller, 2014). Indeed, intended movements can be inferred from neural signals in sensorimotor cortex and converted into control commands of an external device, such as a robotic limb. While remarkable control has been previously achieved, with up to 10 degrees of freedom (Wodlinger et al., 2015), the control of the hand remains relatively primitive, restricted to 4 degrees of freedom. Here, we show that up to 30 degrees of freedom of the upper limb can be simultaneously reconstructed from the responses of a small number of neurons (~20) using a fast and simple decoder.

The dexterity of robotic hands is severely limited by the absence of sensory feedback about hand posture. One approach to convey proprioceptive feedback would be to stimulate proprioceptive neurons in SC (Brian M. London et al., 2008; Salas et al., 2018; Tomlinson & Miller, 2016). Our results suggest that SC neurons – particularly in area 3a – carry a faithful representation of hand posture. However, the topographical organization of this scheme has not yet been established. Indeed, the success of tactile feedback through ICMS has hinged on the somatotopic organization of cutaneous representations in SC. Whether the proprioceptive representation exhibits a spatial topography that can be exploited to convey artificial proprioceptive feedback remains to be elucidated.

Figures

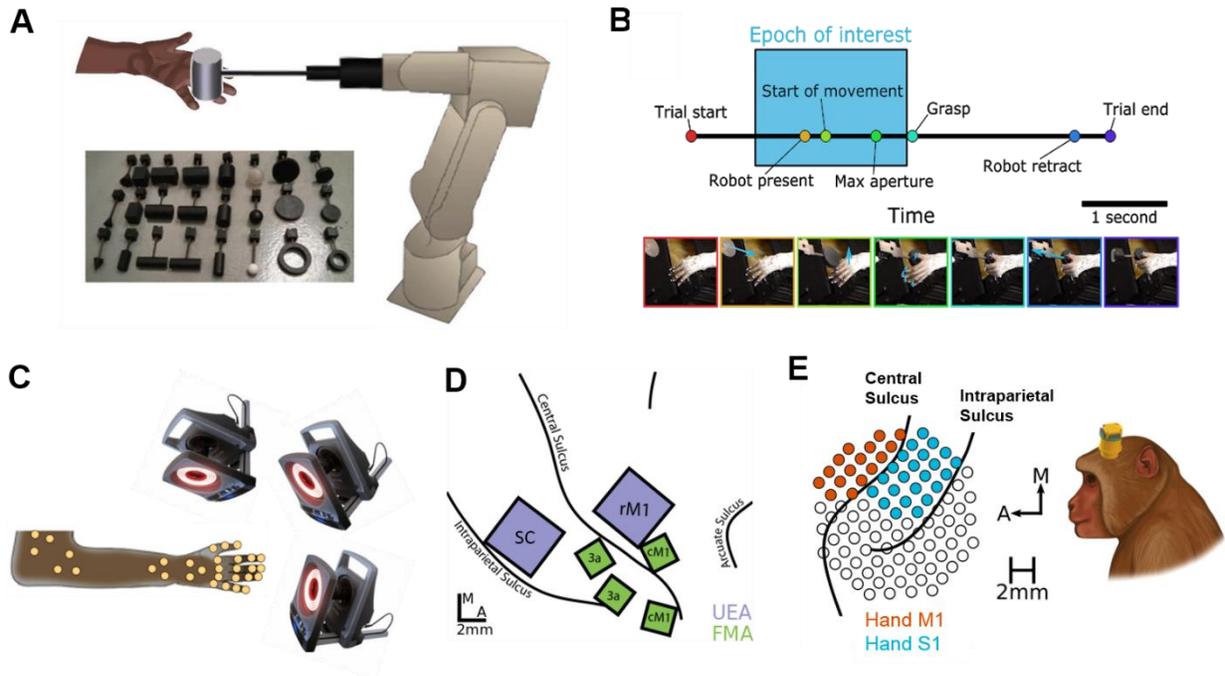

*Figure 1. Behavioral task and experimental set-up. A. On each trial, a robotic arm presented the animal one of 35 objects (inset). B. Trial structure. We focused our analysis on the interval from the start of monkey's limb movement until contact with the object. C. Motion tracking. We placed 30 reflective markers on the animal's joints and tracked their 3D position with a 14-camera Vicon motion tracking system. D. Utah and FMA array placement for monkey 1. E. Coverage of a Grey Matter array (Monkey 3).*

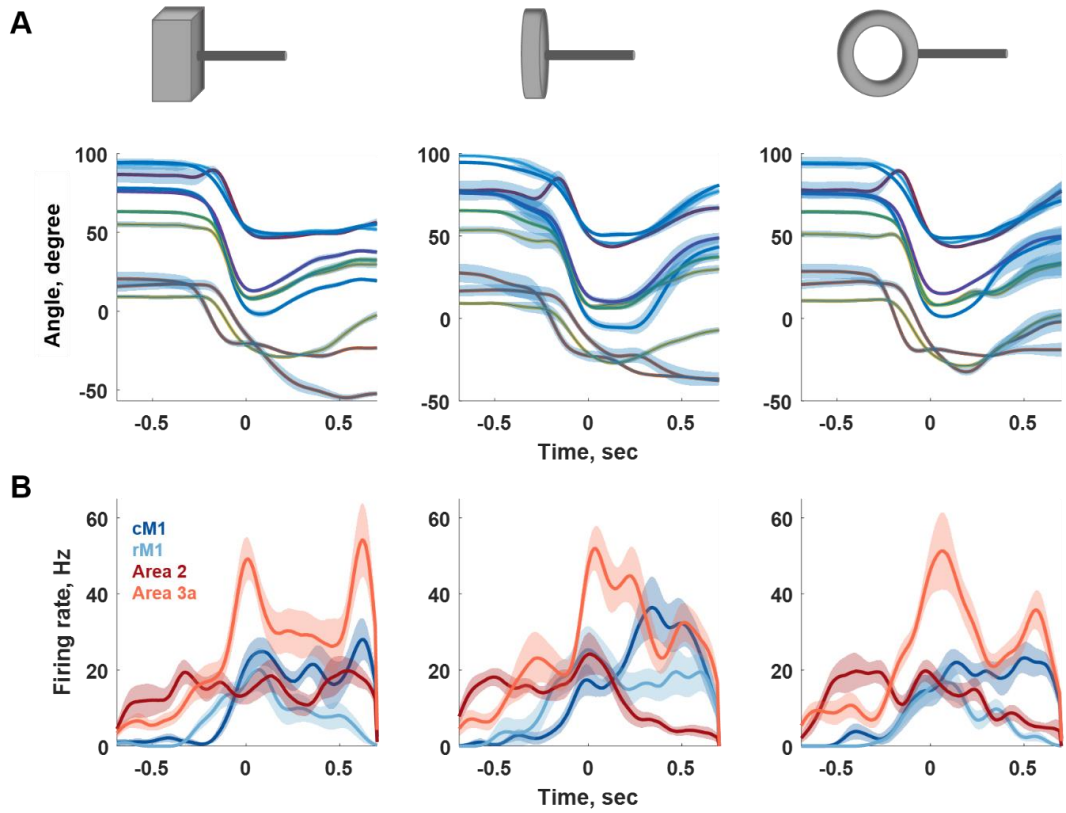

*Figure 2. Example kinematics and neural activity during grasping of three objects. A. Average time-varying angles for 10 joints. The trials were aligned to maximum aperture of the hand (time point 0). B. Example firing rates for four representative neurons from areas cM1 (dark blue), rM1 (light blue), area 2 (red) and area 3a (orange). Shaded regions indicate standard error of the mean.*

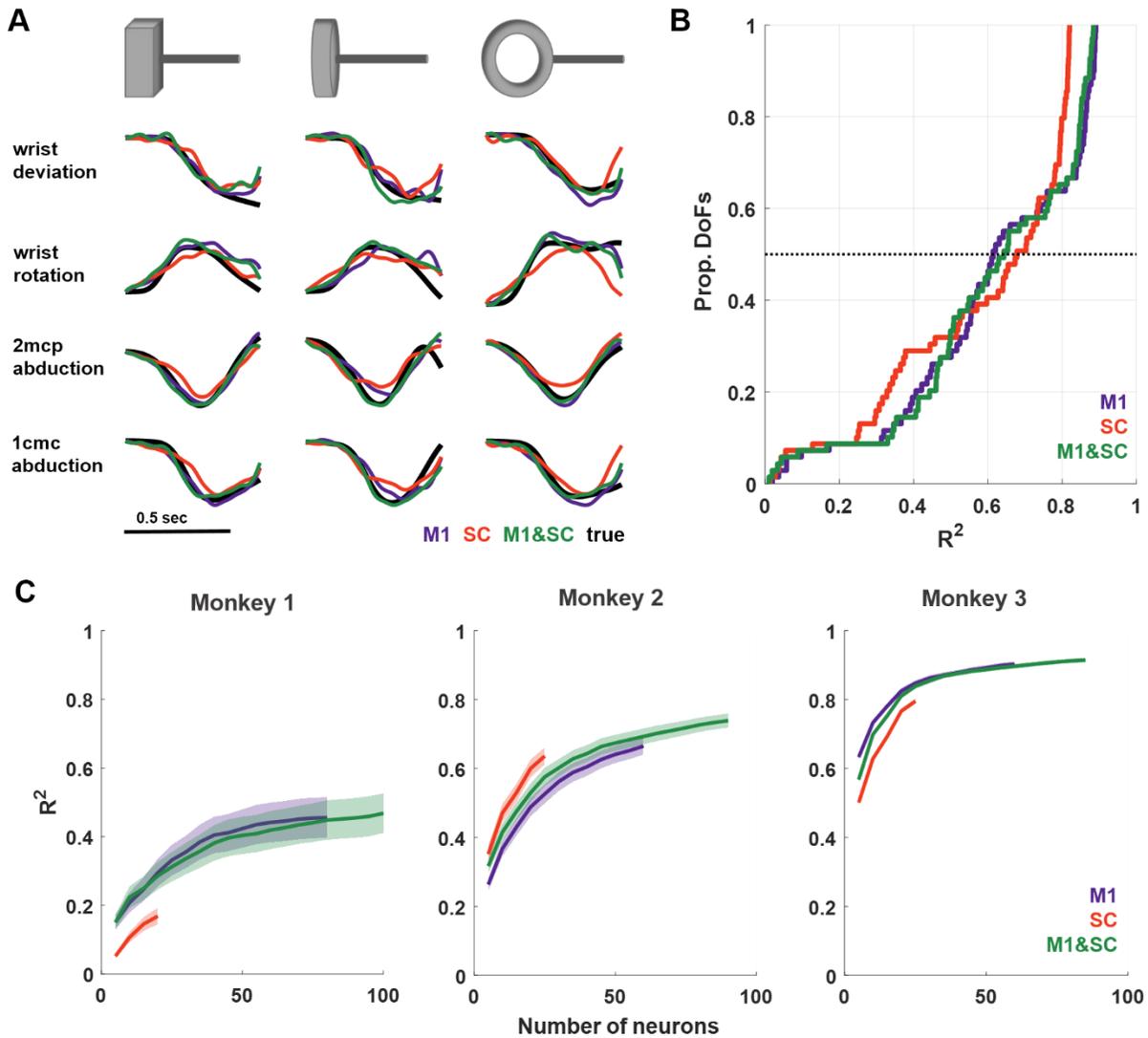

*Figure 3. Decoding kinematics from M1 and SC signals. A. Reconstruction of the time-varying angles of four joints as monkey 3 grasped three objects, using the responses of 20 neurons in M1 (purple), SC (red) and both (green). Black lines indicate measured joint kinematics. B. Cumulative distribution of $R^2$ values for individual joint reconstructions from all monkeys using the responses of 20 neurons in M1 (purple), SC (red), and both (green). C. Performance of decoders as a function of number of neurons in M1 (purple), SC (red), and both (green) for the three monkeys.*

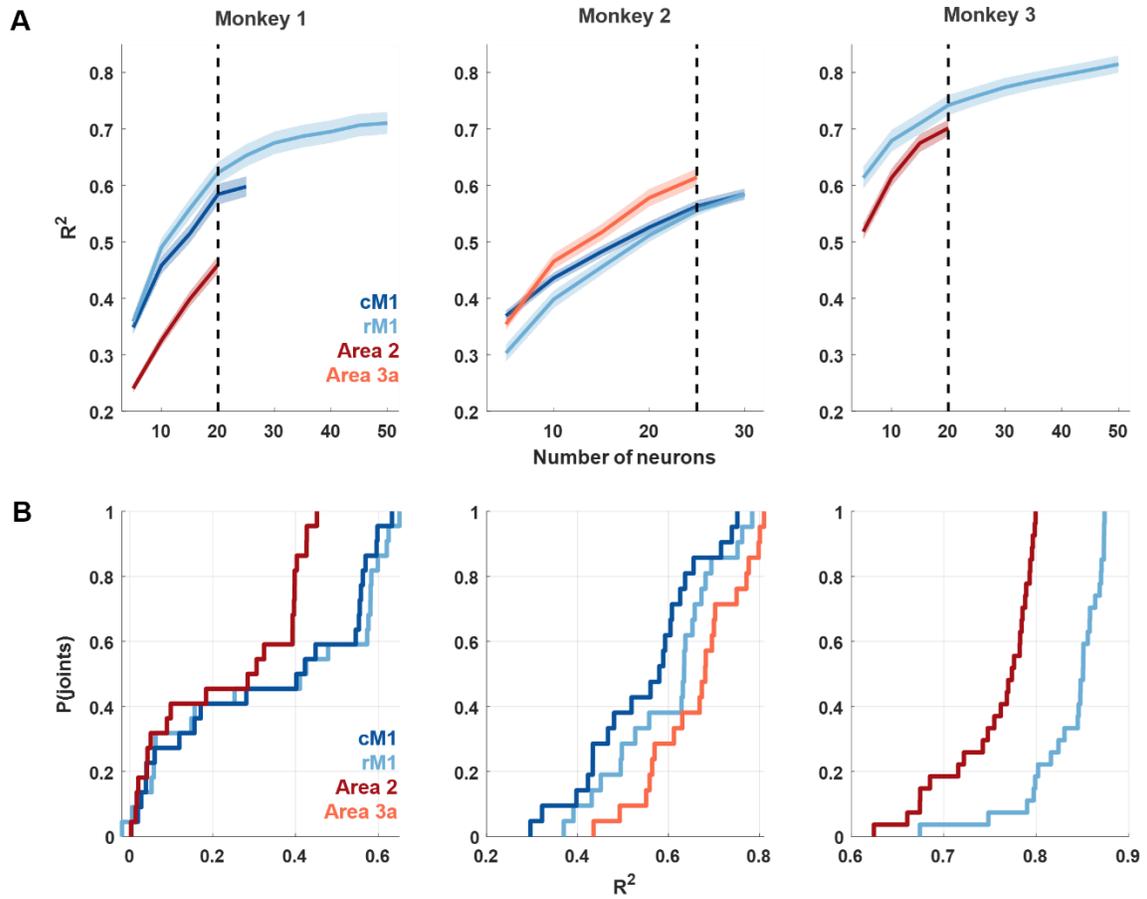

*Figure 4. Comparison of cortical fields. A. Performance as a function of number of neurons for the 10 best joints using cM1 (dark blue), rM1 (light blue), area 2 (red), or area 3a (orange) signals for each of the three monkeys. Thick lines denote the mean and shaded regions the standard errors of (10-fold cross-validated) coefficients of determination. B. Cumulative distribution of $R^2$ for individual joint reconstructions using the same subpopulations as in A, computed for equal numbers of neurons (indicated by a dotted line in subplot A).*

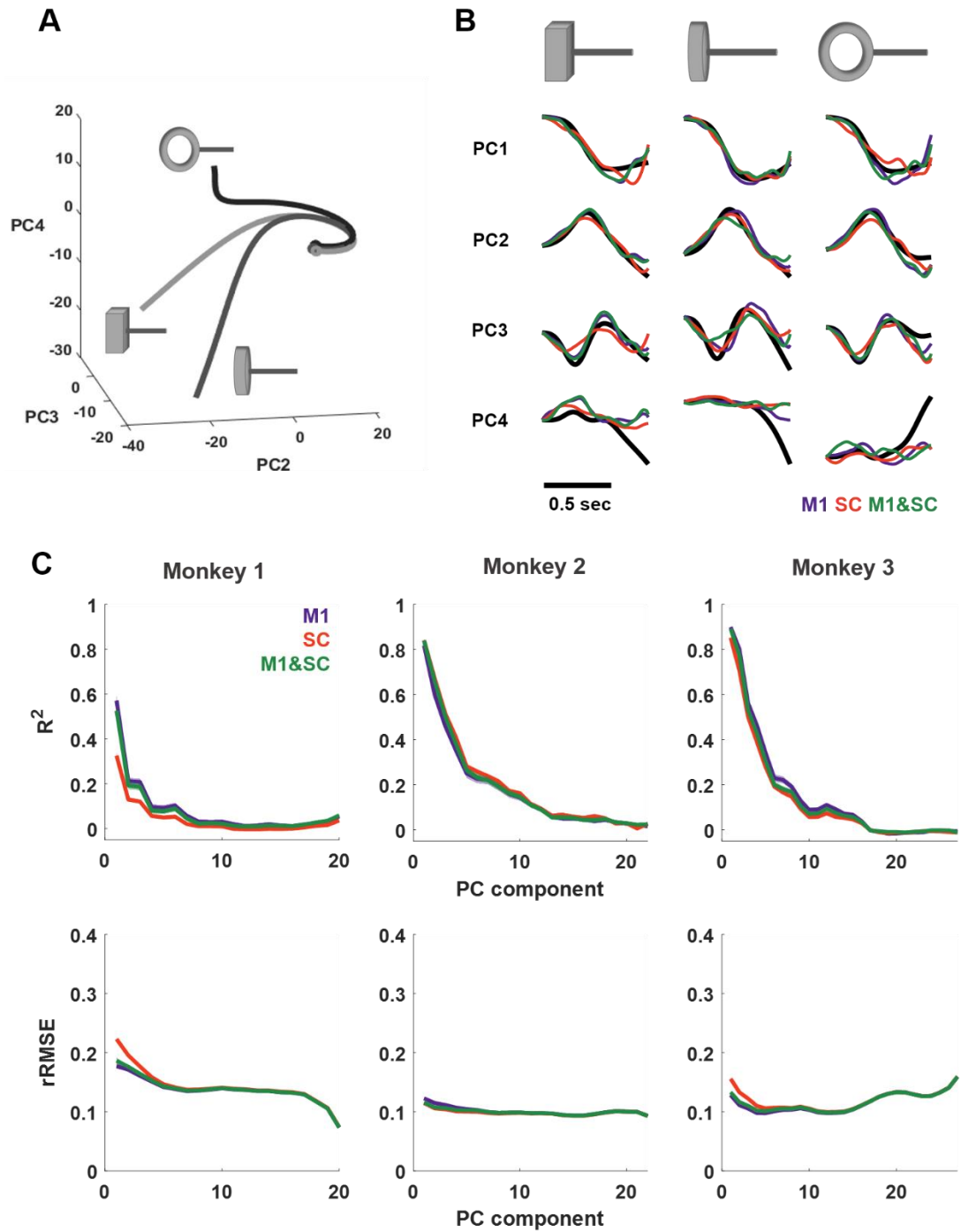

*Figure 5. Decoding synergistic movements. A. Three principal components of the kinematics (PC2-PC4, excluding the condition-independent PC1) of grasping three objects. B. Reconstruction of the first 4 principal components for three objects using the responses of neurons in M1 (purple), SC (orange), or both (green). C. Decoding performance for all principal components ($R^2$ and rRMSE) ranked by the proportion of kinematic variance explained. Shaded region denotes the margin of error computed using 10-fold cross-validation.*

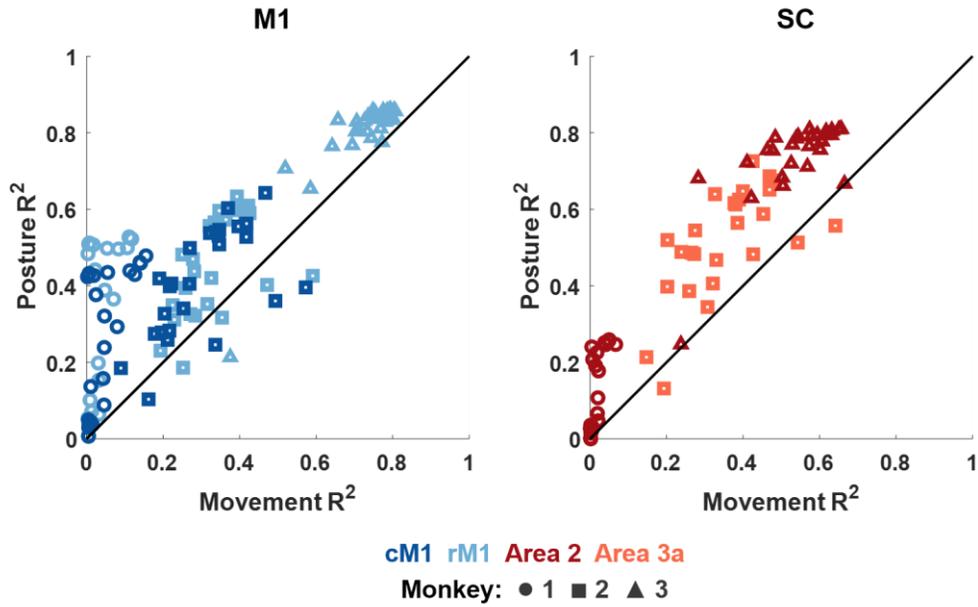

*Figure 6. Posture vs. movement decoding. Decoding performance of posture vs. movement for a randomly selected population of 20 neurons in caudal M1 (dark blue), rostral M1 (light blue), area 2 (red) or area 3a (orange). Each marker denotes the performance for one joint averaged over 10 folds. Different markers denote different monkeys.*

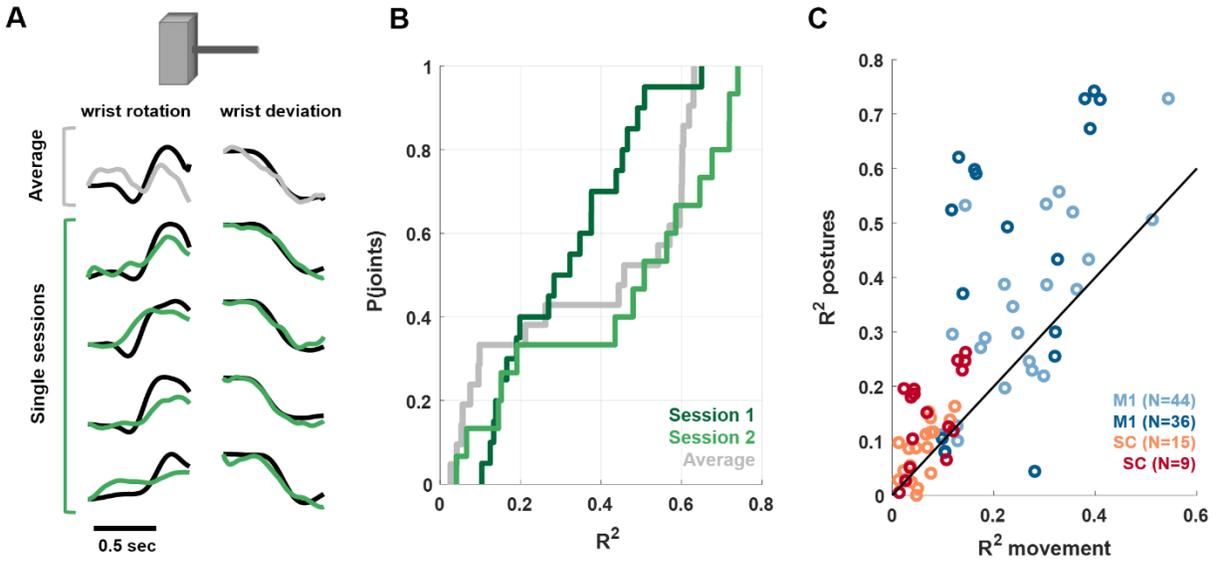

*Figure 7. Single trial decoding. A. Reconstruction of single-trial (green) and mean (gray) kinematics of two wrist degrees of freedom using a population of 44 M1 neurons in monkey 1. B. Cumulative distribution of the coefficient of determination for single and average joint reconstructions using a population of 36 M1 neurons. C. Comparison of movement and posture decoding performance for each reconstructed joint using 2 sessions from monkey 1 with varying numbers of M1 and SC units.*

## Supplementary Figures and Tables

|     |         | Monkey |    |    |
|-----|---------|--------|----|----|
|     |         | 1      | 2  | 3  |
| M1  | cM1     | 28     | 31 | 9  |
|     | rM1     | 52     | 32 | 52 |
|     |         | **80** | **63** | **61** |
| SC  | area 2  | 20     | 1  | 20 |
|     | area 3a | 4      | 28 | 6  |
|     |         | **24** | **29** | **26** |

Table 1. Number of neurons collected across all sessions for each monkey.

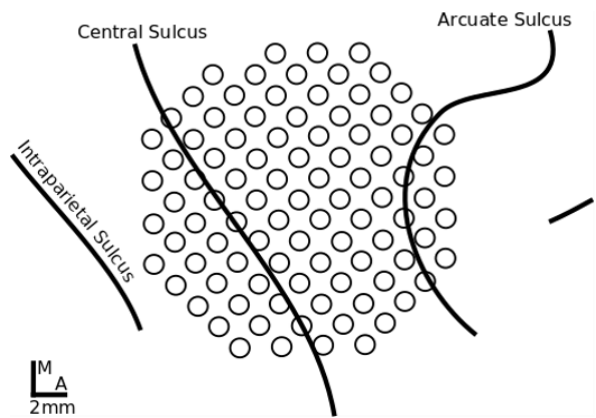

Supplementary Figure 1. Grey matter array placement of monkey 2.

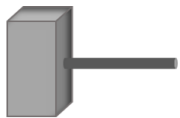
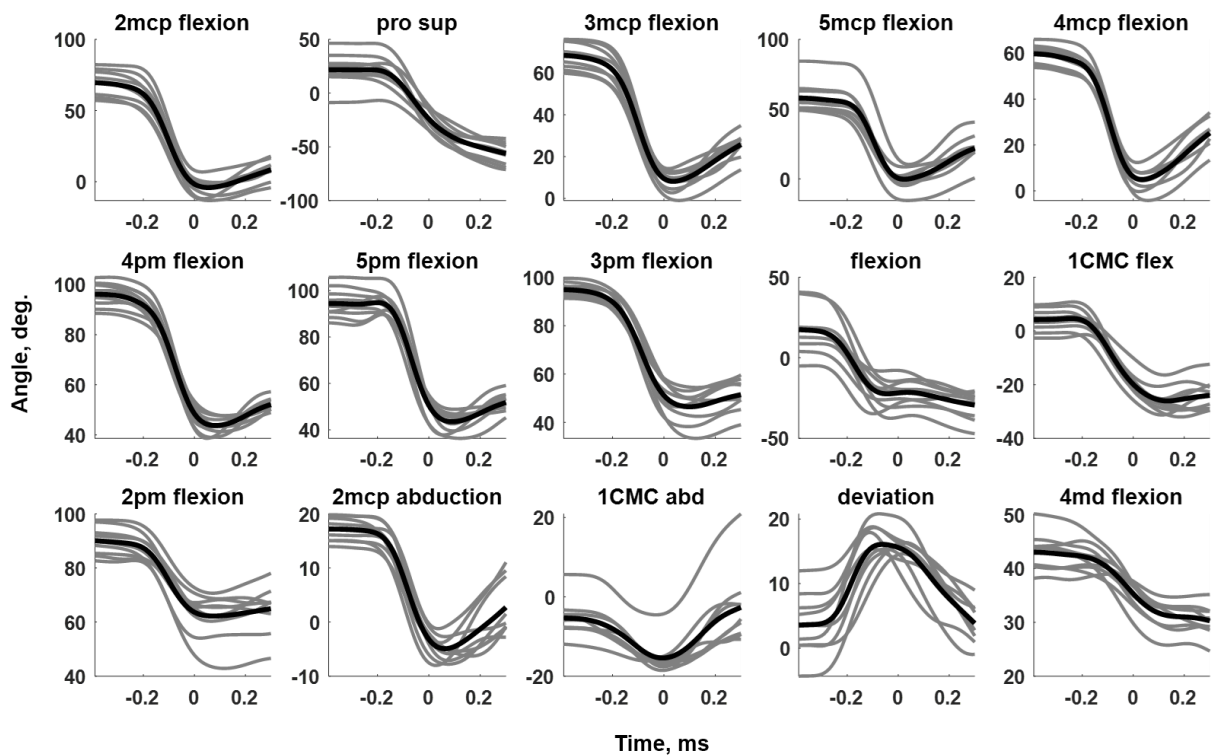

Supplementary Figure 2. Average kinematics within sessions (grey lines) and between sessions (black line) for the 15 highest-variance joints while monkey 3 grasped a large block.

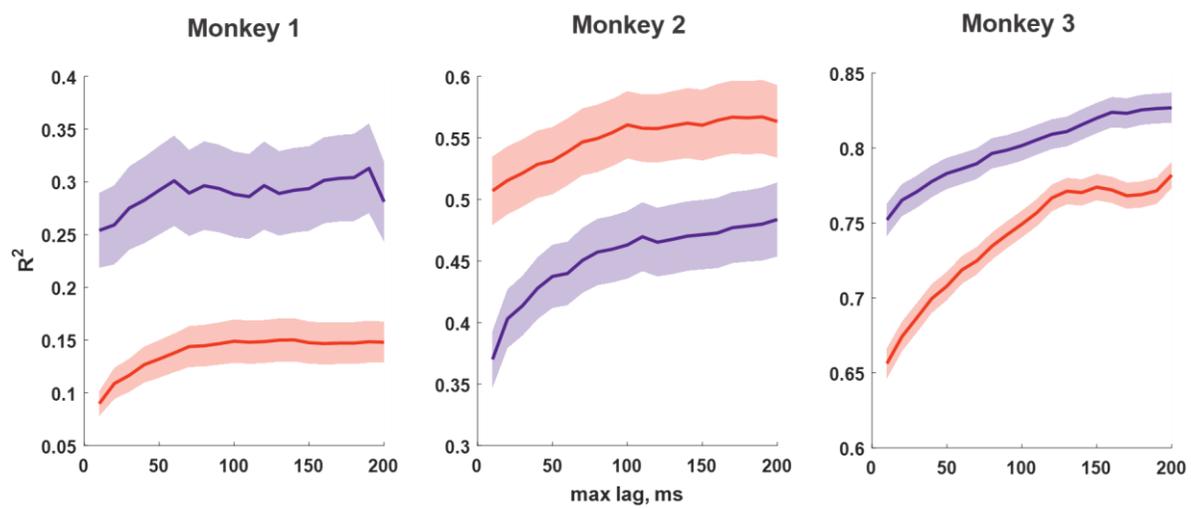

*Supplementary Figure 3. Average performance as a function of lag in neural data for M1 and S1 populations (N=20) for 3 monkeys.*

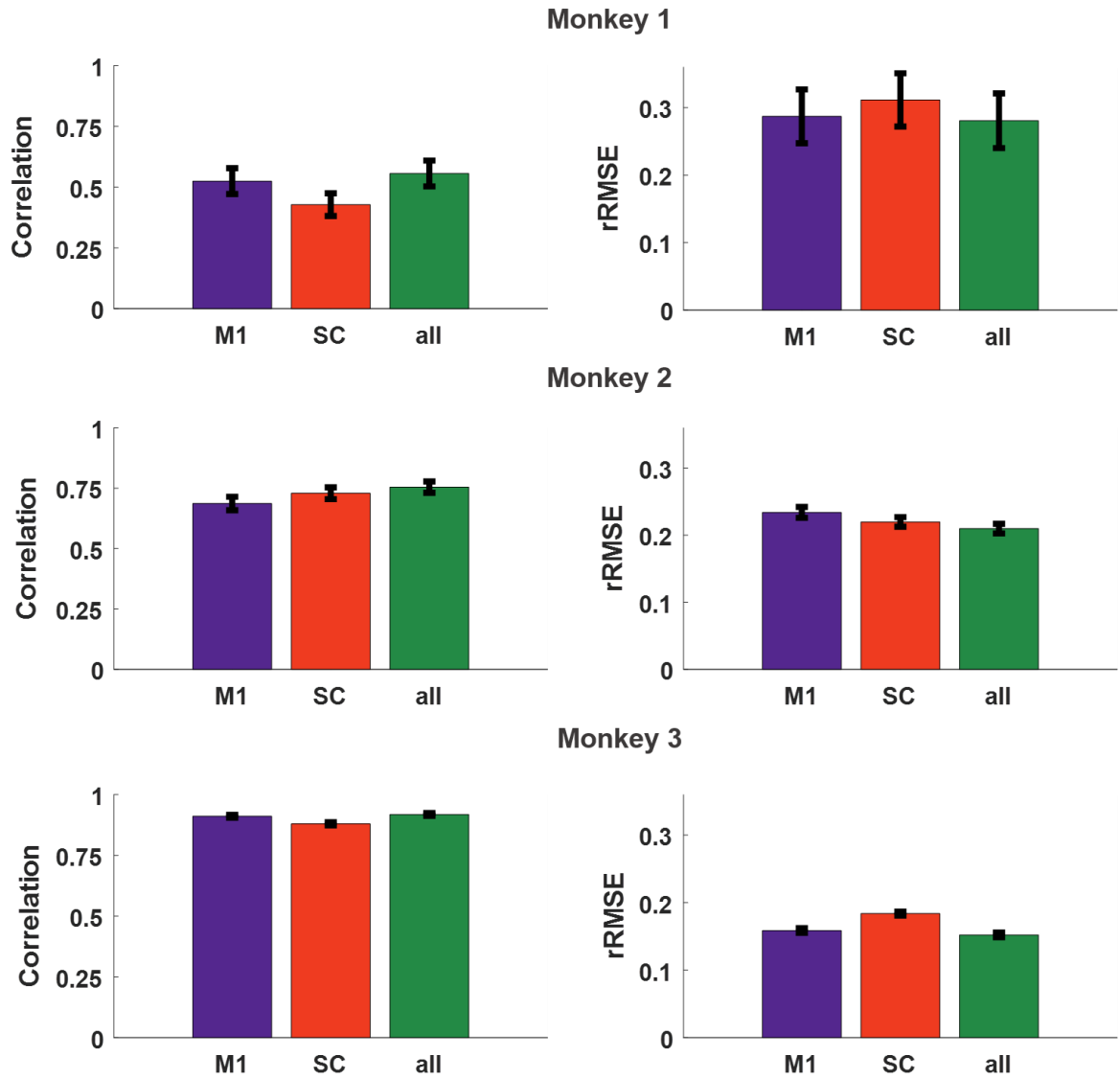

*Supplementary Figure 4. Mean reconstruction performance as gauged by the correlation coefficient (left column) and normalized root mean squared error (right column) for three monkeys. The result is comparable to that reported in Menz et.al. (2015) for M1.*

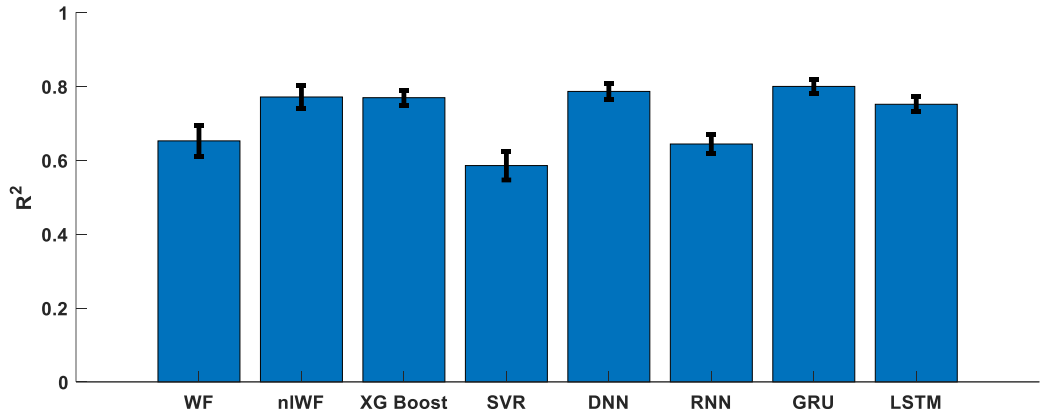

*Supplementary Figure 5. Comparison of a variety of linear and non-linear decoders using M1 responses from monkey 1. WF - Wiener Filter, nlWF - cascade non-linear Wiener filter, XG Boost – Extreme Gradient Boosting, SVR – Support Vector Regression, DNN – dense feedforward neural network, RNN – recursive neural network, GRU – gated recurrent unit, LSTM – Long Short term Memory Network.*

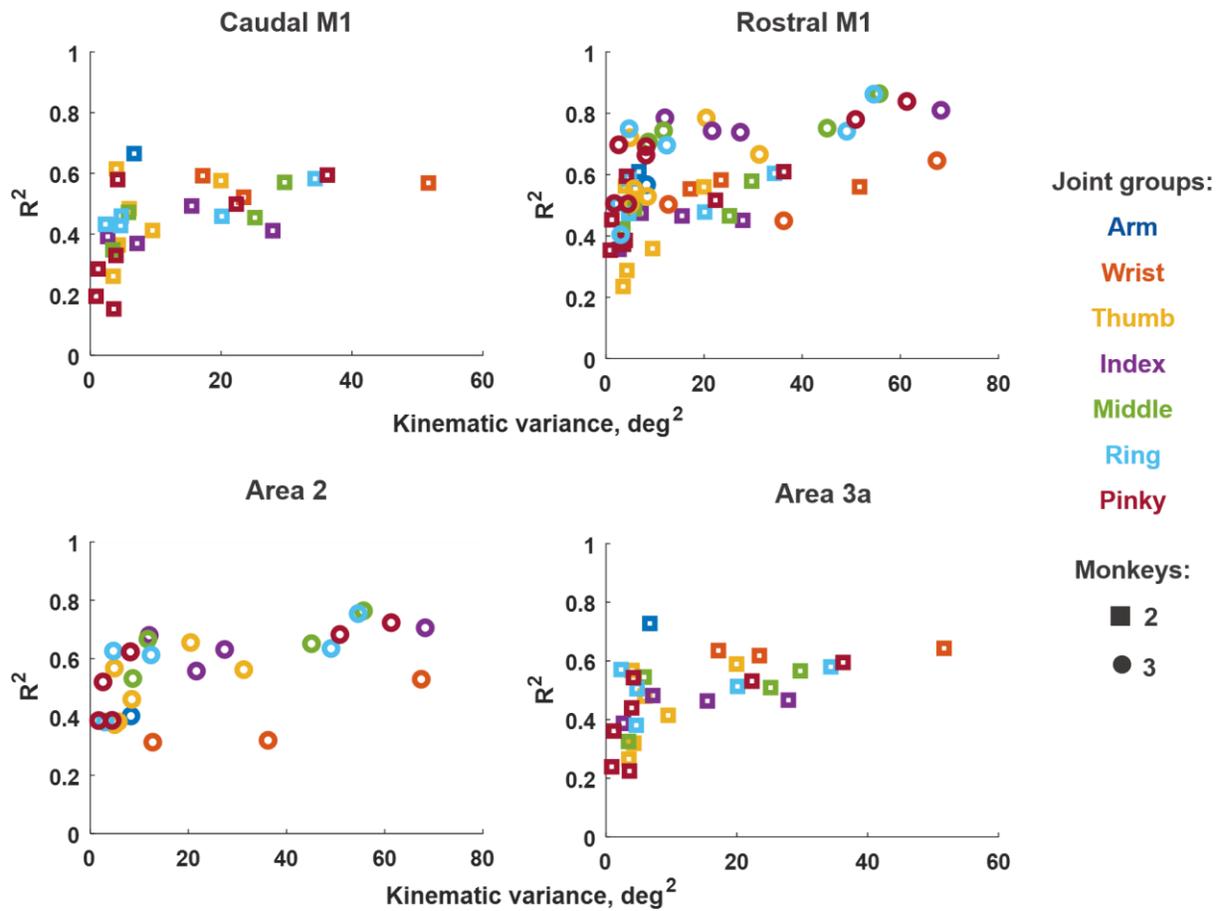

*Supplementary Figure 6. Performance of decoders for single joints, color-coded by joint group: arm (dark blue), wrist (orange), thumb (yellow), index (purple), middle (green), ring (light blue) and pinky (red). Each point denotes the mean over 10-fold cross-validation computed with the maximum number of units in each area. Monkey 1 was omitted from analysis due to an incomplete set of reconstructed joint trajectories.*

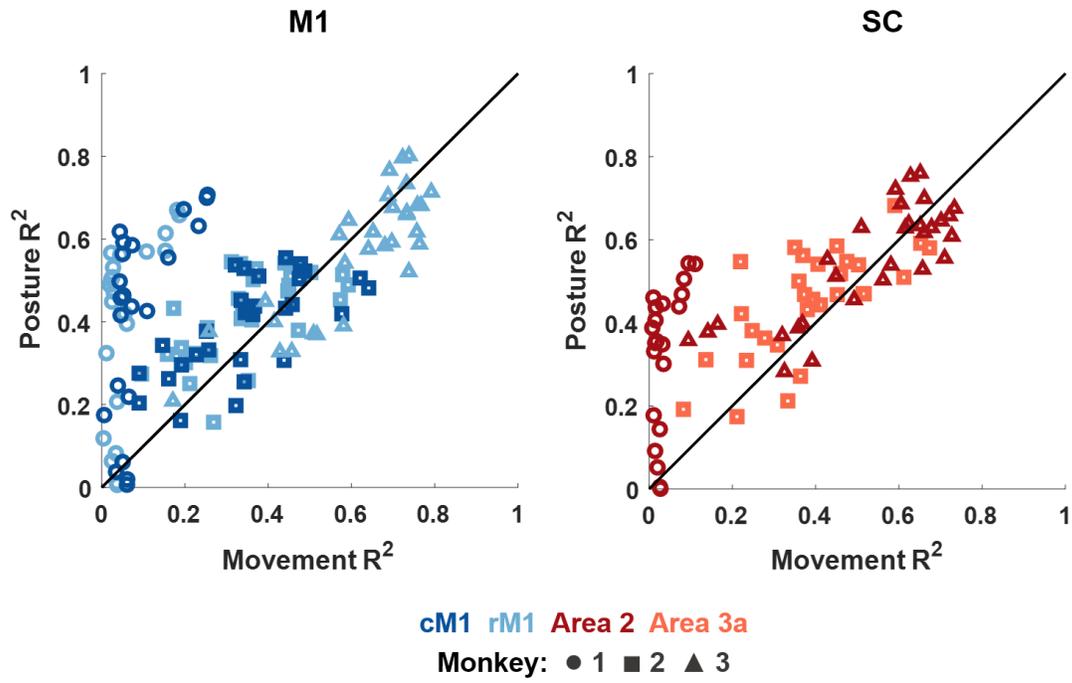

*Supplementary Figure 7. Posture and movement decoding with multiple lags. Decoding performance for posture vs. movement for a randomly selected population of 20 neurons in caudal M1 (dark blue), rostral M1 (light blue), area 2 (red) and area 3a (orange). Each marker denotes the performance for one joint averaged over 10 folds. Different markers denote different monkeys.*